\documentclass[preprint,aps,showpacs,onecolumn]{revtex4} 
\usepackage{amsmath}
\usepackage{amsfonts}
\usepackage{amssymb}
\usepackage{bm}
\usepackage[dvips]{epsfig}
\usepackage[dvips]{graphicx}
\usepackage{float}
\usepackage{dcolumn}
\usepackage{ifpdf}
\usepackage{dcolumn}
\usepackage{multirow}

\newcommand{\be}{\begin{equation}}
\newcommand{\ee}{\end{equation}}
\newcommand{\bes}{\begin{subequations}}
\newcommand{\ees}{\end{subequations}}
\newcommand{\ben}{\begin{eqnarray}}
\newcommand{\een}{\end{eqnarray}}

\begin{document}
\title{False vacuum decay in kink scattering}
 \author{Adalto R. Gomes$^{1}$, F. C. Simas$^{2}$, K. Z. Nobrega$^{3}$, P. P. Avelino$^{4,5,6}$}
 \email{argomes.ufma@gmail.com, simasfc@gmail.com, bzuza1@yahoo.com.br, pedro.avelino@astro.up.pt}
  \noaffiliation
\affiliation{
$^1$ Departamento de F\'\i sica, Universidade Federal do Maranh\~ao (UFMA)
Campus Universit\'ario do Bacanga, 65085-580, S\~ao Lu\'\i s, Maranh\~ao, Brazil\\
$^2$ Centro de Ci\^encias Agr\'arias e Ambientais-CCAA, Universidade Federal do Maranh\~ao
(UFMA), 65500-000, Chapadinha, Maranh\~ao, Brazil\\
$^3$ Departamento de Eletro-Eletr\^onica, Instituto Federal de Educa\c c\~ao, Ci\^encia e
Tecnologia do Maranh\~ao (IFMA), Campus Monte Castelo, 65030-005, S\~ao Lu\'is, Maranh\~ao, Brazil\\
$^4$ Instituto de Astrof\'{\i}sica e Ci\^encias do Espa{\c c}o, Universidade do Porto, CAUP, Rua das Estrelas, PT4150-762 Porto, Portugal\\
$^5$ Centro de Astrof\'{\i}sica da Universidade do Porto, Rua das Estrelas, PT4150-762 Porto, Portugal\\
$^6$ Departamento de F\'{\i}sica e Astronomia, Faculdade de Ci\^encias, Universidade do Porto, Rua do Campo Alegre 687, PT4169-007 Porto, Portugal
}
\noaffiliation

\begin{abstract}
In this work we consider kink-antikink and antikink-kink collisions in a modified $\phi^4$ model with a false vacuum characterized by a dimensionless parameter $\epsilon$. The usual $\phi^4$ model is recovered for $\epsilon=0$. We investigate the $\epsilon<<1$ regime where the kink in the presence of false vacuum can be understood as a small deformation of the standard kink for the $\phi^4$ model. We show that the attractive interaction between the kink-antikink pair leads to a rich scattering pattern, in some cases delaying considerably the false vacuum decay.
\end{abstract}

\pacs{ 04.60.Kz, 11.27.+d}

\keywords{kink, lower dimensional models, extended classical solutions}

\maketitle


\section{ Introduction }

False vacua  play an important role in fundamental physics, in particular in the context of  inflationary models \cite{infl1,infl2,infl3}, electroweak phase transition and baryogenesis \cite{electroweak1,electroweak2,electroweak3,electroweak4,electroweak5}, bubble collisions \cite{bubble1,bubble2,bubble3,bubble3a,bubble4,bubble4a,bubble5}, signals for the possible metastability of Higgs vacuum \cite{sm-vacuum1,sm-vacuum2,sm-vacuum3} and dark energy models \cite{dark-energy,Avelino1}.
Soliton solutions in nonlinear field theory  describe localized energy concentrations which are able to propagate without  modifying their shape.  They have large applicability in condensed matter physics \cite{bisc}, optics \cite{KiAg}, quantum field theory \cite{masu}, nuclear physics \cite{nucl1,nucl2,nucl3} and cosmology \cite{vile,agjo,Avelino2,Avelino3}. The simplest soliton solutions are the kink and antikink in $(1,1)$-dimensions.

Kink scattering processes in nonintegrable models have intrincate structures. This has been studied not only in the simple $\phi^4$ model \cite{sug,moshir,csw,w1,campbell,belkud,aom,gh,saada,doreyrom}, but also in potentials of higher self-interaction \cite{domerosh,dedeke,gankudli,weigel,roman,begani1,galeli,begani2} and non-polynomial potentials \cite{peycam,gankud,gaaes,bazbegani1,bazbegani2,sgn,siadnool}. Further interesting directions of investigation include the interaction of a kink with an impurity \cite{gooha,feikiva,goohol}, kinks in models of two scalar fields \cite{harosh,alonso1,alonso2,vakiru}, multi-kink collisions \cite{mgsdj,masd,sadmke}, boundary scattering \cite{dhmr, adp} and models with generalized dynamics \cite{twin}.

Embedded in higher dimensional spaces, kinks give rise to domain walls and branes \cite{Avelino4}. In some cyclic universe scenarios, domain walls were considered as a possibility to generate the Big Bang conditions \cite{khovsttu,yutama,paan,branes}, where planar symmetry is assumed and the dynamics is effectively $(1,1)$ dimensional.
Solitons are also studied in the context of bubble collisions in the early universe. In the limit of a high nucleation rate per unit four-volume in comparison to $H^4$ (where $H$ is the Hubble expansion rate), one can neglect the spacial curvature of the universe and consider the bubbles as in flat spacetime \cite{bubbles}. In this regime, collisions of two bubbles have $SO(2,1)$ symmetry and are described by a $(1,1)$ dimensional wave equation.

 The mechanism of false vacuum decay in scalar field theories is well understood \cite{decay1,decay2}. In particular, for an asymmetric double well potential, domain wall collisions with planar and $SO(2,1)$ symmetry were considered in Ref. \cite{braden1}. Domain wall collision with such high degree of symmetry can be described as collisions between a kink and an antikink in $(1,1)$ dimensions. In this background the effects of small initial planar \cite{braden1} and nonplanar \cite{braden2} small fluctuations were investigated. However, Ref. \cite{braden1} only provided some examples for the output of background collisions, with no description of their dependence either with initial velocity or with the parameter controlling the  difference in potential energy between the two wells.

 The present work intends to fill this gap, with the aim of investigating the kink scattering and the corresponding false vacuum evolution, considering the situation where the false vacuum initially occupies the restricted region between the kink-antikink pair and that where the false vacuum occupies the unbounded region outside the area defined by the antikink-kink system.  We consider symmetric kink-antikink and antikink-kink collisions, with the pair initially at equal distances from the center of mass and initial velocities with same modulus. We show how the attractive interaction between the kink and antikink is the main factor affecting the structure of the scattering and, in some cases, delaying considerably the complete decay of the false vacuum even in a restricted region around the center of mass of the kink-antikink pair.

In the next section we review some of the main properties of the $\phi^4$ model. The numerical results are presented in the Sect. III, and we conclude in the Sect. IV.


 \section{The model}

We consider the action in $(1,1)$-dimensions in Minkowski spacetime with a Lagrangian with standard dynamics
\be
S=\int dtdx\biggl(\frac12\partial_\mu\phi\partial^\mu\phi-V(\phi)\biggr)
\ee
and a modified $\phi^4$ potential
\be
V(\phi)=\frac\lambda4\bigg(\phi^2-\frac{M^2}\lambda \bigg)^2-\frac12\epsilon\frac{M^3}{\sqrt\lambda} \phi,
\label{potential}
\ee
where $\epsilon$ is a dimensionless self-coupling parameter. From here on we will consider $\epsilon\geq0$ without loosing generality. For $\epsilon>0$ the potential has a local minimum at $\phi=-1$ (the false vacuum), a local maximum at $\phi=0$ and a global minimum at $\phi=1$ (the true vacuum).  Here we are interested in the $|\epsilon|\ll1$ limit, where the presence of the linear term in Eq. (\ref{potential}) leads to a small force on the $\phi^4$ kink that favors the expansion of the true vacuum region \cite{kish}. For sake of definiteness, we take $\lambda=M^2=2$. In this case equation of motion is given by
\be
\label{eom}
\ddot\phi-\phi''-2\phi+2\phi^3-\epsilon=0
\ee
Linear stability analysis around the static kink solution with $\phi(x,t)=\phi_K(x)+ \cos(\omega t)\eta(x,t)$ leads to a Schr\"odinger-like equation:
\be
-\eta''+(-2+6\phi_K^2)\eta=\omega^2\eta.
\ee
Static solutions can be found for $\epsilon=0$, given by the usual $\phi^4$ kink $\phi_K(x)=\tanh(x)$ and antikink $\phi_{\bar K}(x)=-\tanh(x)$. These solutions, of minimal energy, can also be obtained by first-order BPS equations \cite{bps1,bps2}. Stability analysis of the $\phi^4$ kink have eigenvalues (and corresponding eigenfunctions)  \cite{sug}: $\omega_0^2=0$ (the zero-mode or translational mode) and $\omega_1^2=3$ (the vibrational mode), followed by a continuum of states with $\omega_k^2=4+k^2$. The existence of vibrational modes is important for the comprehension of the complex behavior of kink-antikink scattering. One such effect is known as two-bounces, and it is characterized by a collision process where the translational energy is stored in the kink-antikink pair during an amount of time. As a result, the pair, after being scattered, oscillates around the contact point and retrocede for a second collision. This mechanism was described in the Ref. \cite{csw} as an exchange of energy between the translational and vibrational modes. Despite applicable in the present case, we remark there are some known exceptions to this mechanism: in the $\phi^6$ model, two-bounces occur  even in the absence of a vibrational mode \cite{domerosh};  the presence of more than one vibrational mode can result in the suppression of two-bounce windows \cite{siadnool}. Also, quasinormal modes can store energy during a collision in the $\phi^4$ model \cite{doreyrom}. Now back to our problem for $\epsilon\neq0$, as far as we know, there are no explicit solutions of the equation of motion. In particular, corrections $\phi_n$ to the kink field in $n$th order in $\epsilon$ were presented in Ref. \cite{kish}, and
are described in terms of the basis of eigenfunctions $\{ \eta_n \}$.


\section { Numerical Results }


The kink scattering processes have a strong dependence on whether the initial region with false vacuum is finite or not. Then in the following we will consider separately kink-antikink and antikink-kink collisions.


\subsection{False vacuum in an unbounded interval: kink-antikink collisions}


First of all let us consider a symmetric kink-antinkink collision. We take as initial conditions a $\phi^4$ kink with velocity $v$ and a $\phi^4$ antikink with velocity $-v$:
\begin{eqnarray}
\phi(x,0) &=& \phi_{K}(x+x_0,v,0)-\phi_{K}(x-x_0,-v,0)-1, \\
\dot\phi(x,0) &=& \dot\phi_{K}(x+x_0,v,0)-\dot\phi_{{K}}(x-x_0,-v,0). 
\end{eqnarray}
We fixed $x_0=15$ as the initial kink position, i.e., the kink solution centered at $-x_0$ and the antikink at $x_0$. Note that the initial condition is only a solution for $\epsilon=0$. We apply this even for $\epsilon\ll1$ considering that the system will relax by emitting scalar radiation during their free propagation toward the collision region. To solve Eq. (\ref{eom}) we use a pseudospectral method on a grid with $2048$ nodes with periodic boundary conditions for $\phi$ and $\dot\phi$ and we set the grid boundary at $x_{max}=200$.  A sympletic method with the Dirichlet condition imposed at $x=\pm x_{max}$ was also applied to double check our numerical results. We used a $4^{th}$ order finite-difference method with spatial step $\delta x=0.09$ and a $6^{th}$ order symplectic integrator with time step $\delta t=0.04$.

\begin{figure}
\includegraphics[{angle=0,width=5cm,height=5cm}]{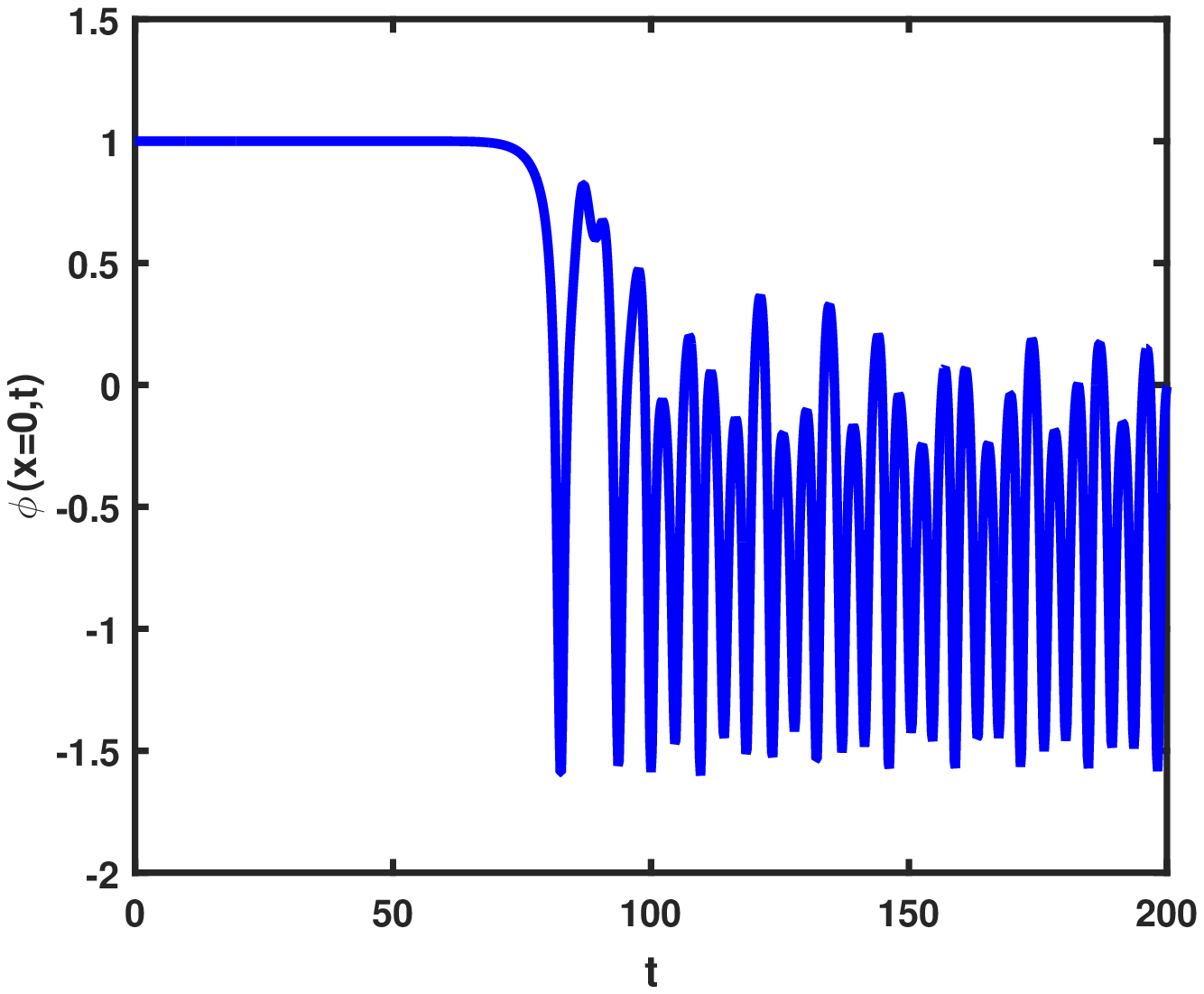}
\includegraphics[{angle=0,width=5cm,height=5cm}]{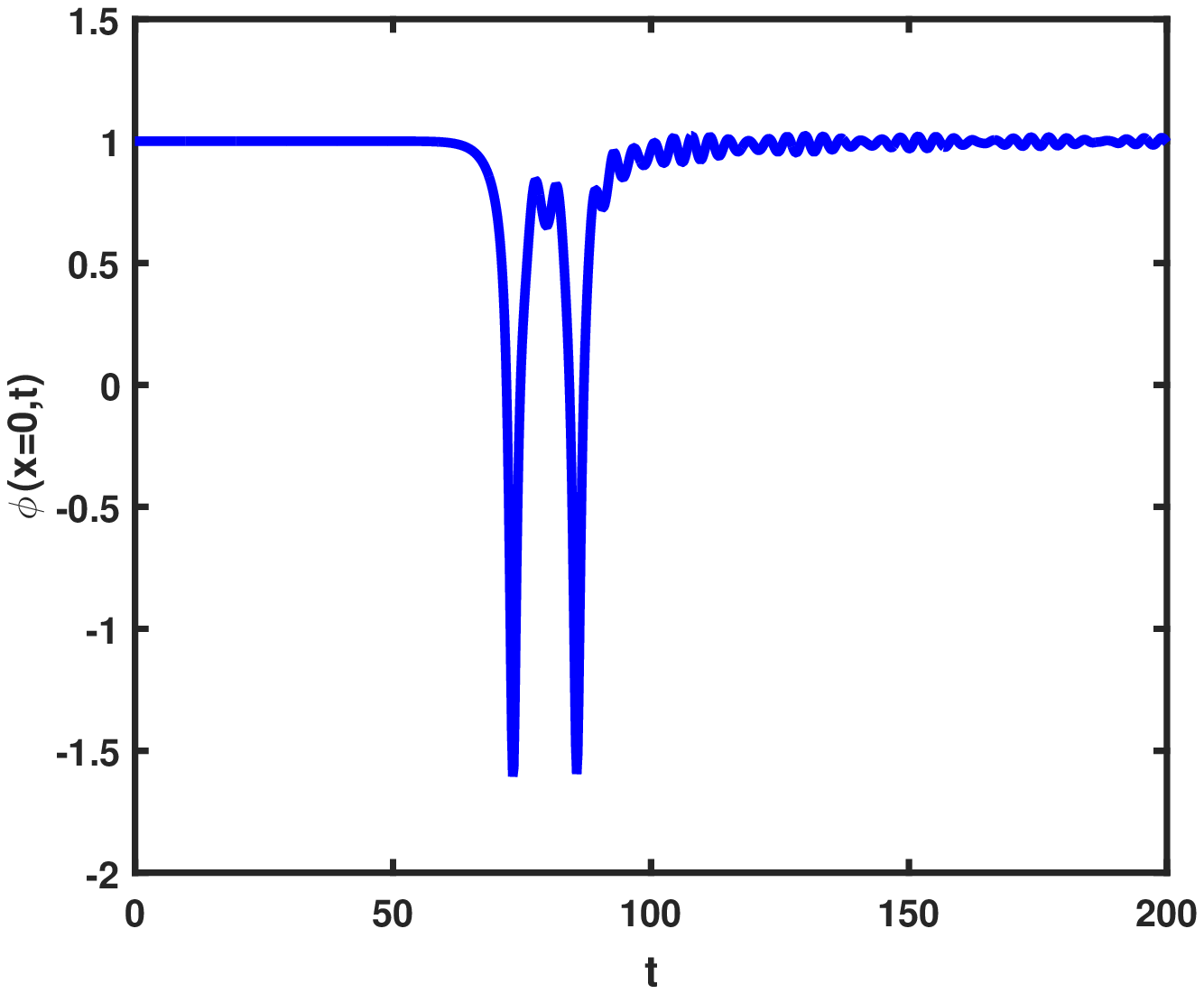}
\includegraphics[{angle=0,width=5cm,height=5cm}]{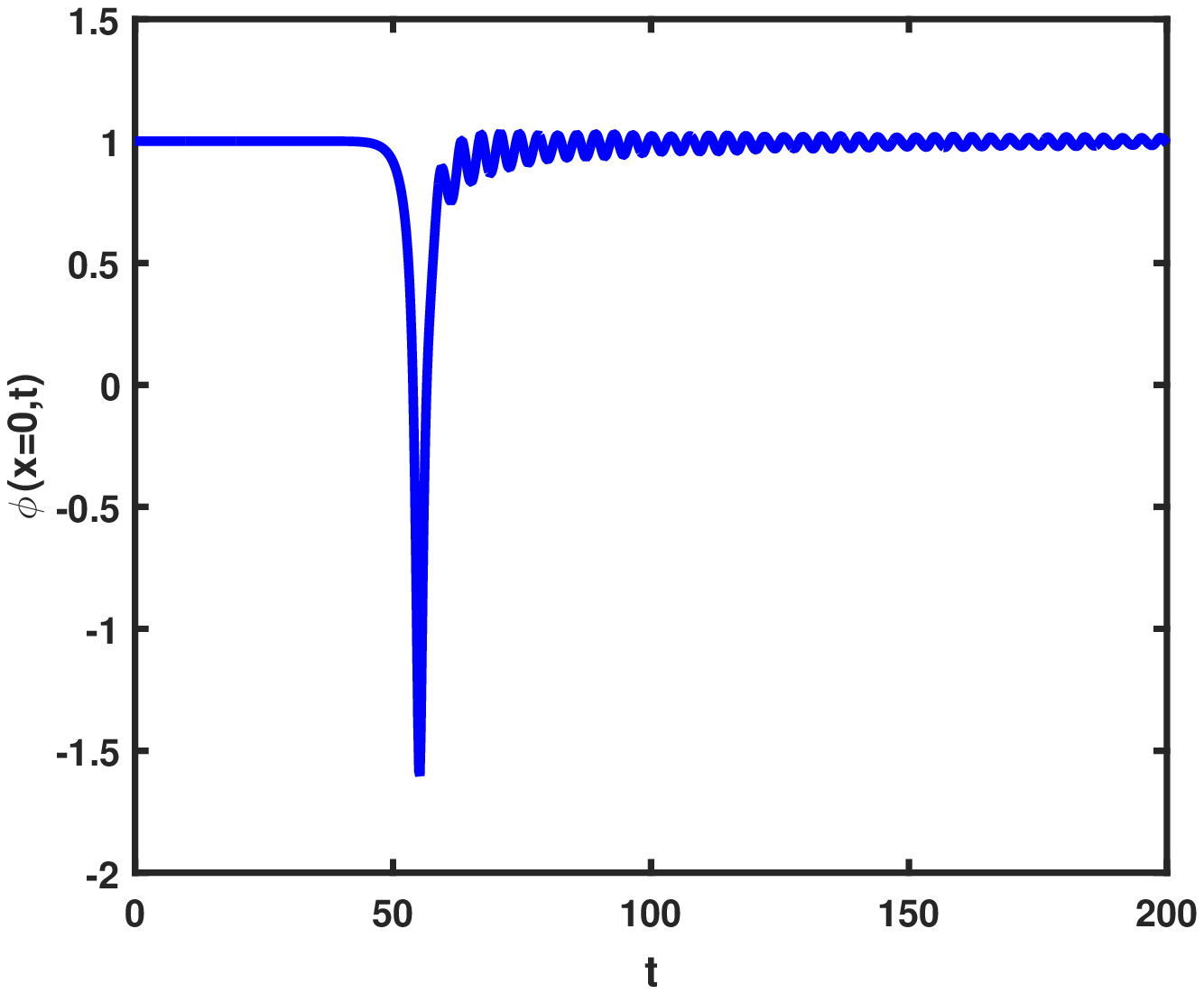}
\caption{False vacuum in an unbounded interval: kink-antikink collisions. Scalar field at the center of mass, $\phi(0,t)$, showing a) the formation of bion state (for $v=0.17$), b) two-bounce collision (for $v=0.1926$) and c) one-bounce collision (for $v=0.26$). Results are for the $\phi^4$ model, corresponding to $\epsilon=0$.}
\label{fig-phi0}
\end{figure}

We start our analysis by reviewing some aspects of the $\epsilon=0$ case, corresponding to the $\phi^4$ theory, and well described in the literature \cite{sug,moshir,csw,w1,campbell,belkud,aom,gh,saada,doreyrom}. Figs. \ref{fig-phi0}a-\ref{fig-phi0}c depict three profiles of scalar field at the center of mass as a function of time, $\phi(0,t)$, for different values of the initial velocity. Initially the point $x=0$ is at the vacuum $\phi=1$.
For small velocities, the output of the scattering process is a bion state - see an example in Fig. \ref{fig-phi0}a - where the pair radiates with the scalar field oscillating without a recognizable pattern until, in the long run, finishing at the vacuum $\phi=-1$. Velocities above a critical velocity $v_{crit}\sim 0.26$ show an inelastic scattering, where the output field retains the initial value $\phi=1$ after one bounce, as in the example of Fig. \ref{fig-phi0}c. We could characterize this inelastic scattering as having $N_b=1$ bounces. For smaller velocities with $v\lesssim v_{crit}$ there are some  velocity windows where the scalar field presents two-bounces (then $N_b=2$) during the collision process, as shown in Fig. \ref{fig-phi0}b.

\begin{figure}
\includegraphics[{angle=0,width=8cm,height=6cm}]{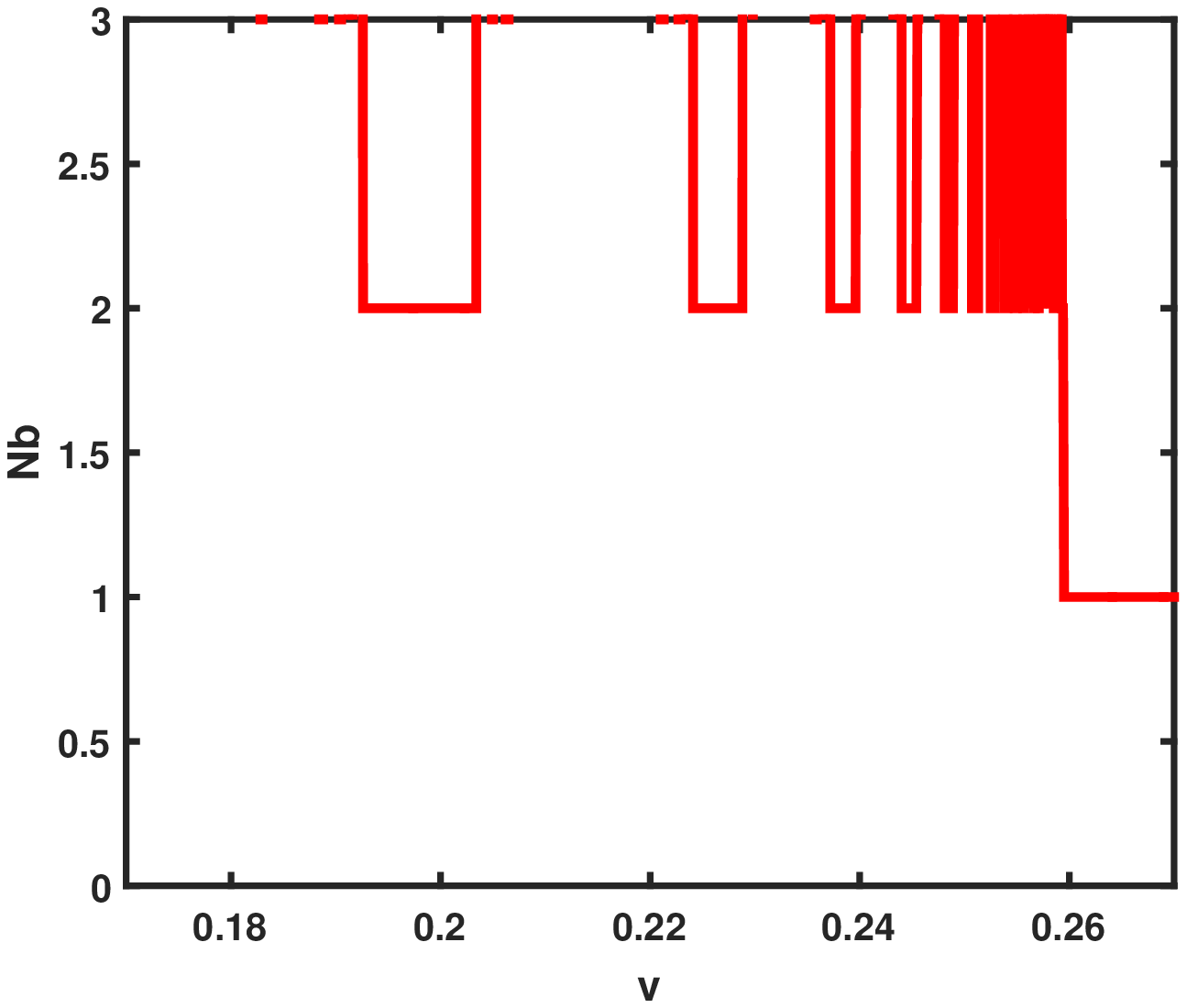} 
\includegraphics[{angle=0,width=8cm,height=6cm}]{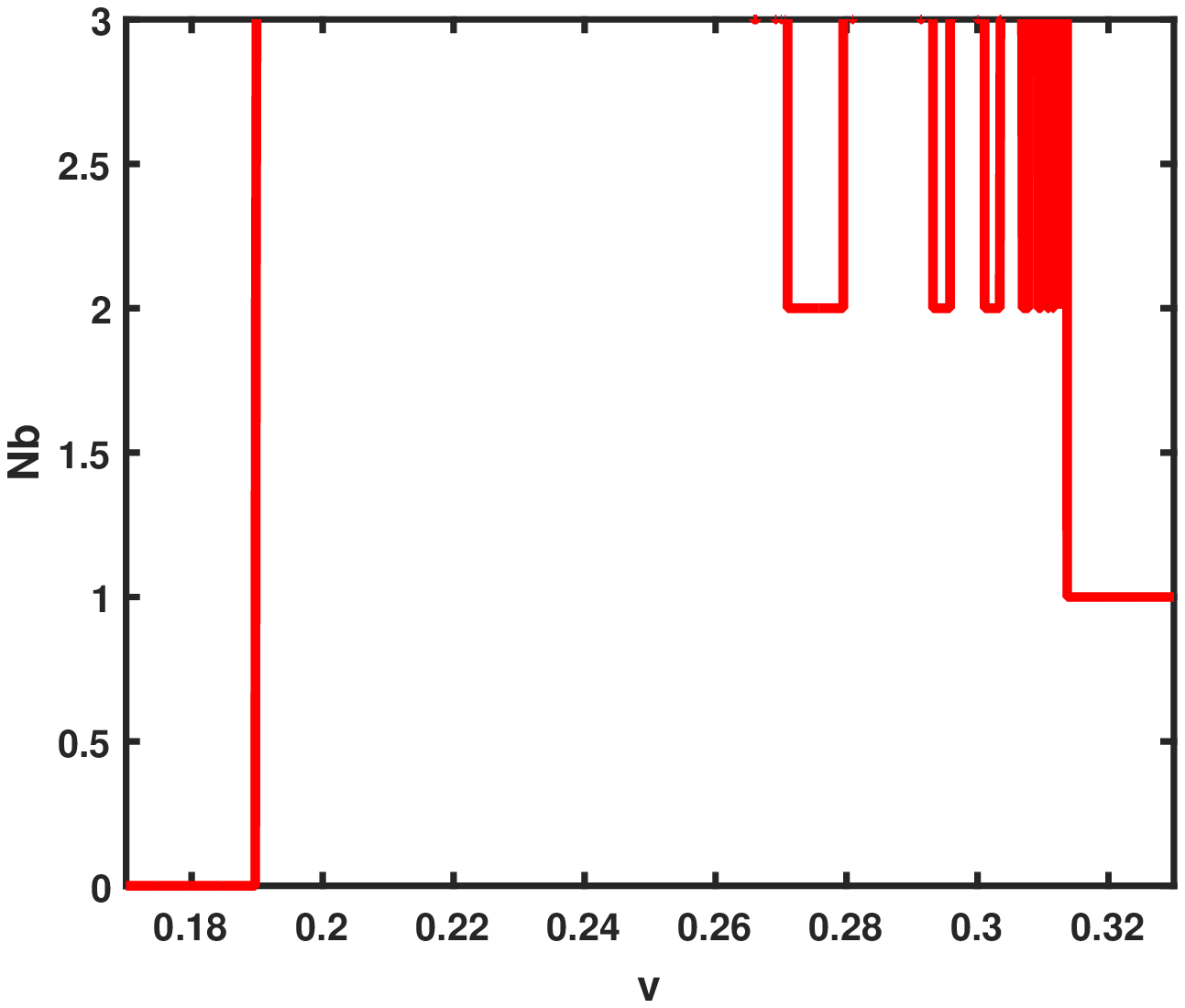} 
\includegraphics[{angle=0,width=8cm,height=6cm}]{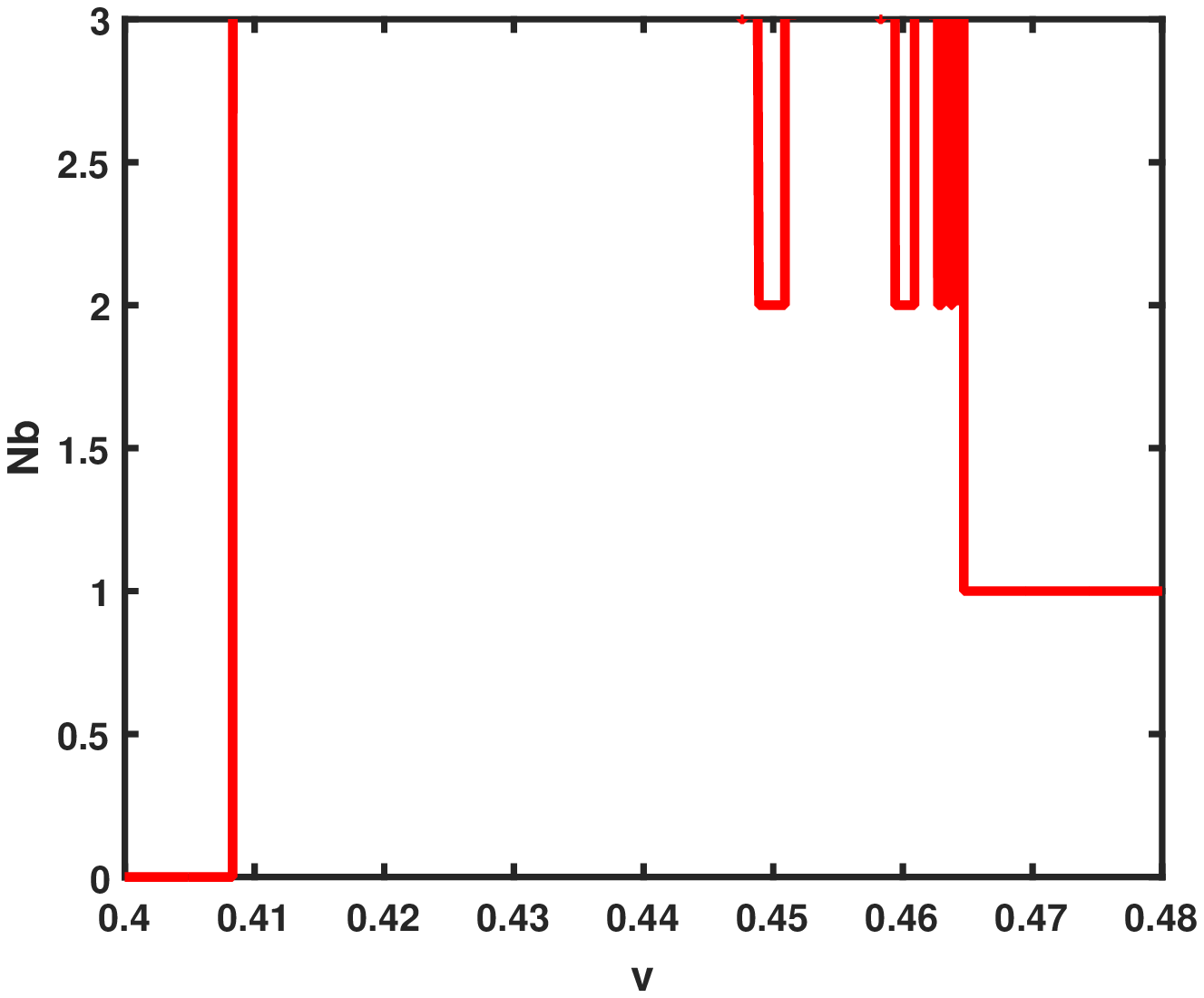} 
\includegraphics[{angle=0,width=8cm,height=6cm}]{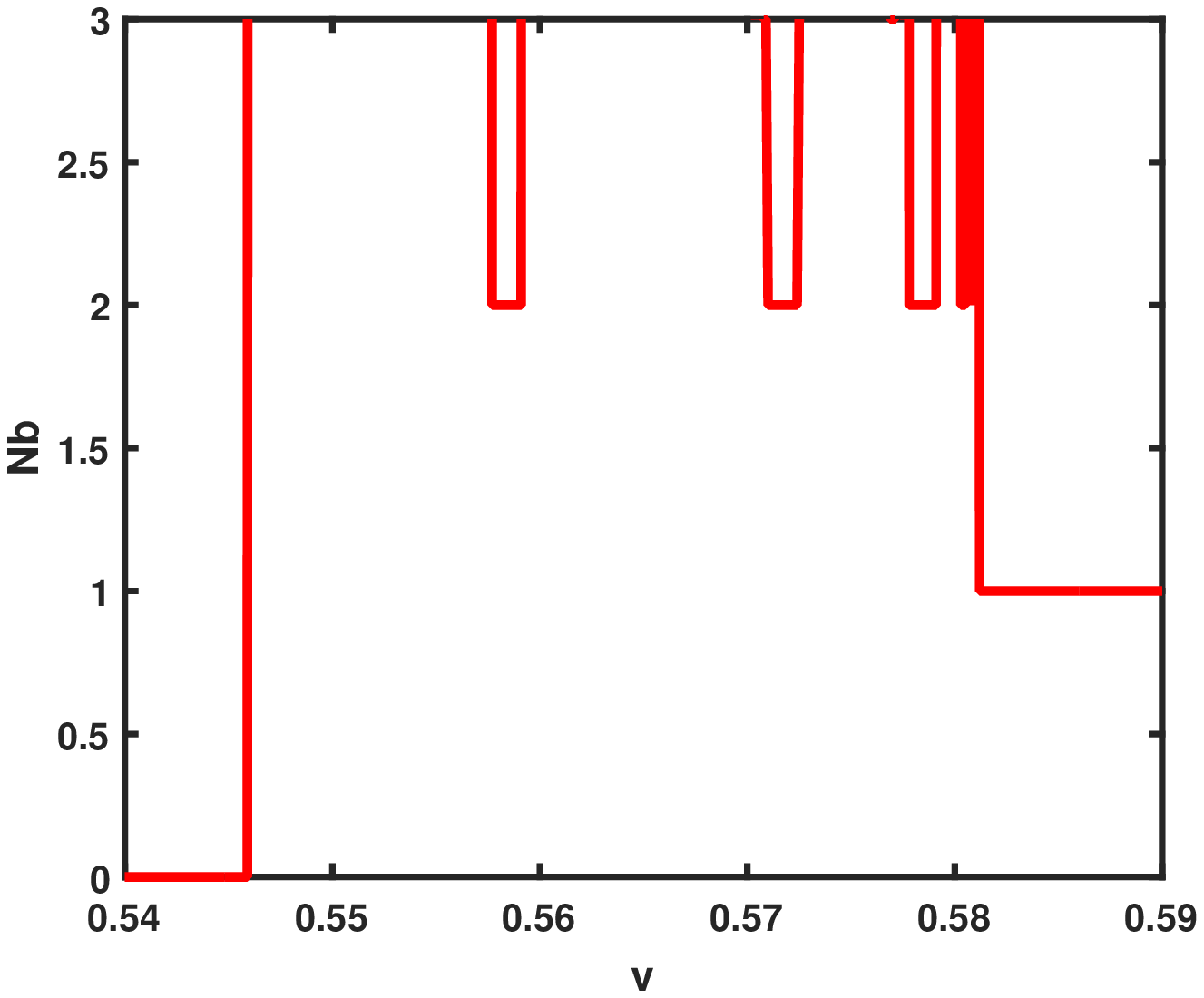} 
\caption{False vacuum in an unbounded interval: kink-antikink collisions. Number of bounces versus initial velocity $v$ for (a) $\epsilon=0$, (b) $\epsilon=0.001$, (c) $\epsilon=0.005$ and (d) $\epsilon=0.01$.}
\label{fig_Nbxv}
\end{figure}

The structure of such two-bounce windows is better visualized in Fig. \ref{fig_Nbxv}a, which shows the number of bounces as a function of the initial velocity. As shown in Fig. \ref{fig-phi0}a, bion states have a large number of oscillations of $\phi(x,0)$. Large values of $N_b$ are not directly represented in Fig. \ref{fig_Nbxv}, but there one represent bion states as making frontier with the two-bounce windows. Each two-bounce window is labeled by an integer $m$, the number of oscillations of $\phi(0,t)$ between the bounces. For instance, in Fig. \ref{fig-phi0}b $m=1$, meaning that this collision belong to the first two-bounce window from Fig. \ref{fig_Nbxv}a.

\begin{figure}
\includegraphics[{angle=0,width=5cm,height=5cm}]{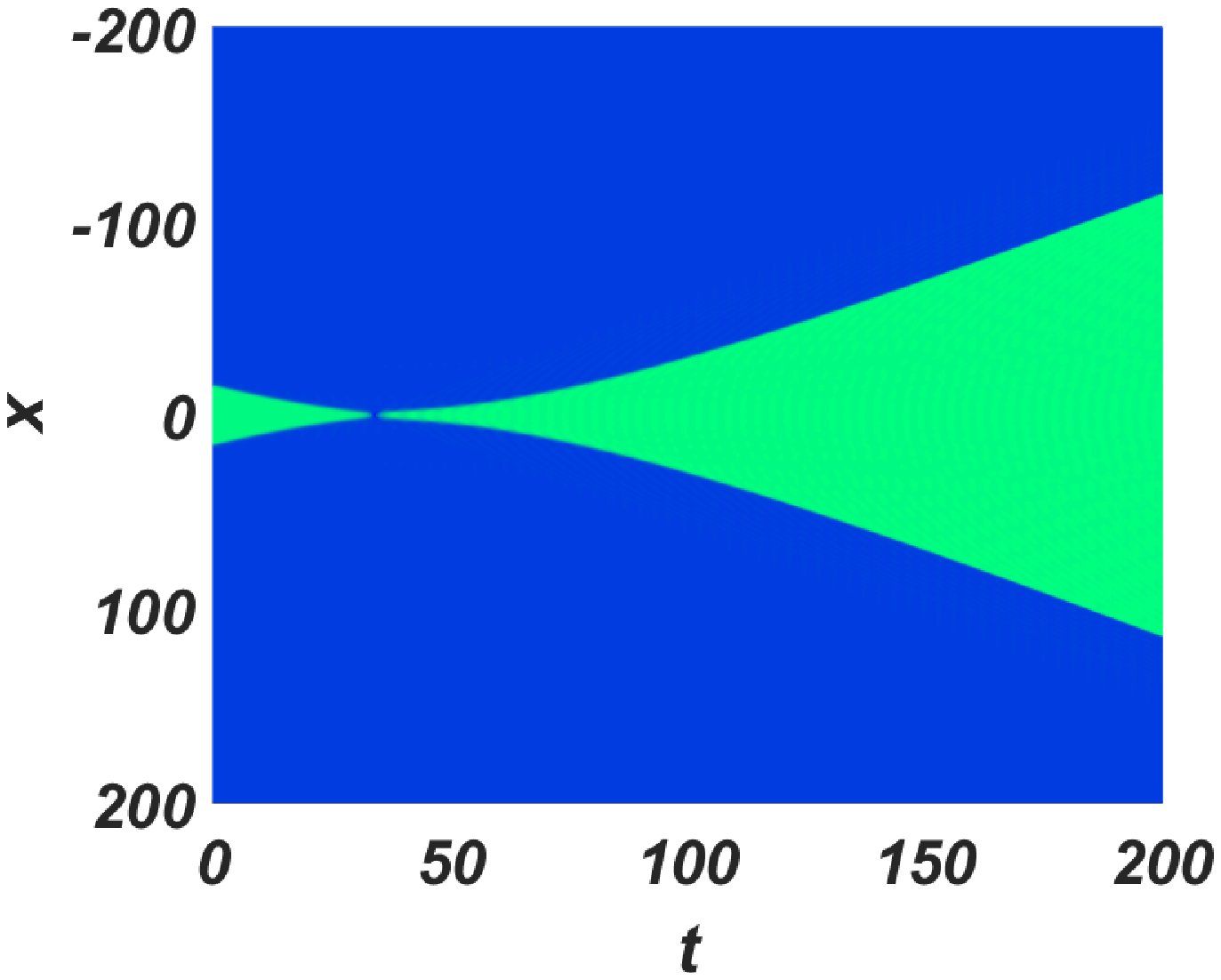} 
\includegraphics[{angle=0,width=5cm,height=5cm}]{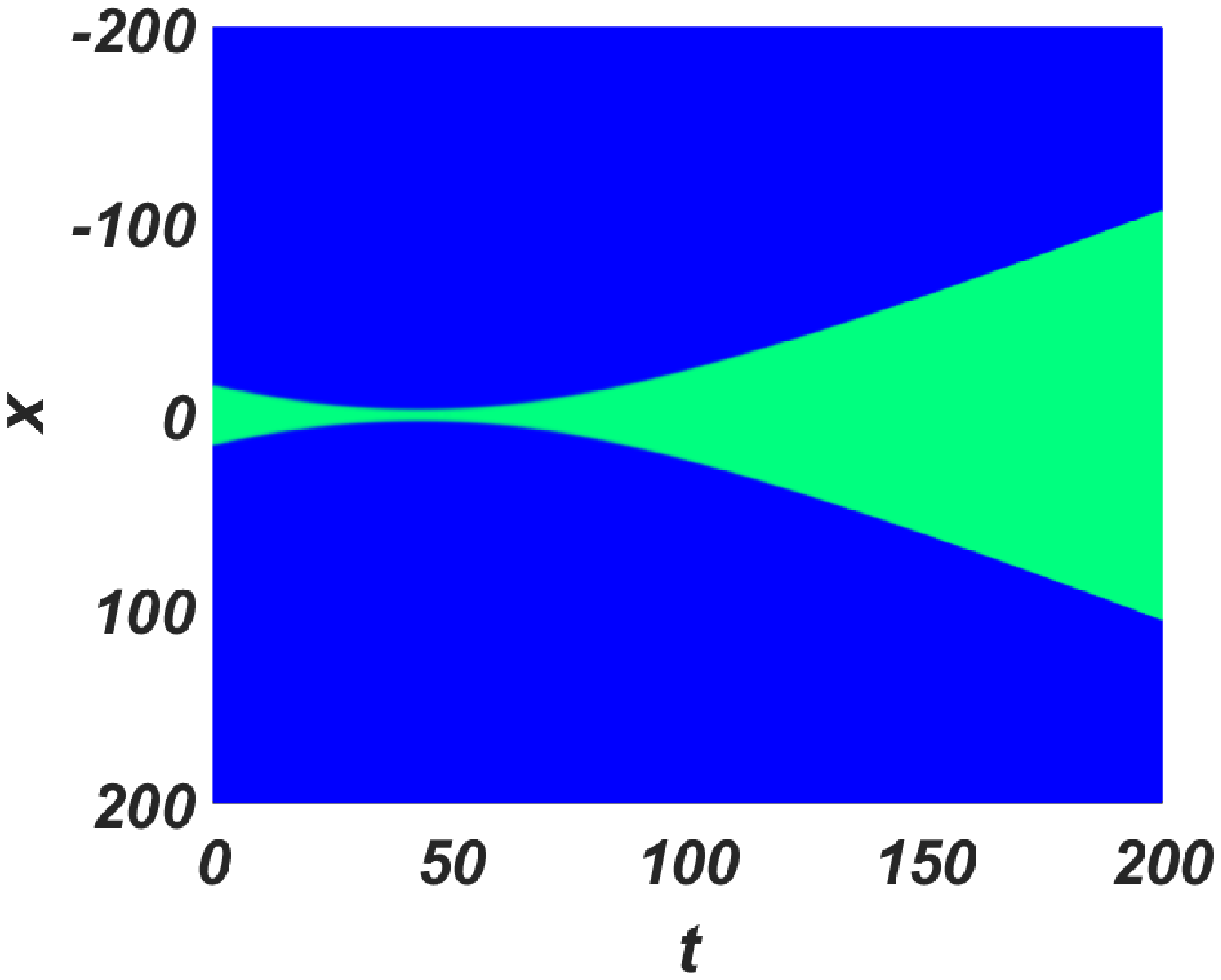} 
\includegraphics[{angle=0,width=5cm,height=5cm}]{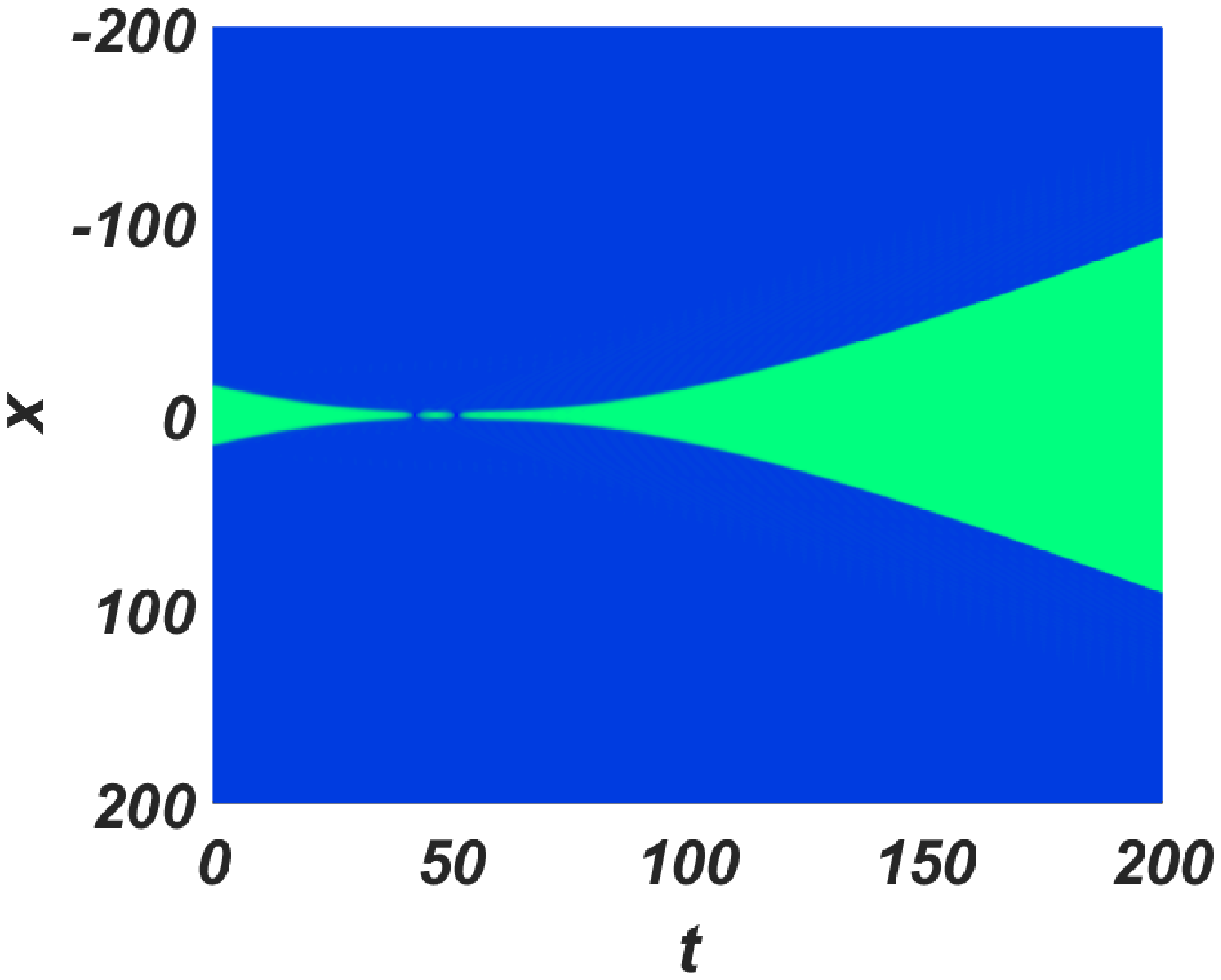} 
\caption{False vacuum in an unbounded interval: kink-antikink collision for $\epsilon=0.01$ with (a) $v=0.59$, (b) $v=0.54$ and (c) $v=0.558$. Compare with diagram of Fig.\ref{fig_Nbxv}d. Blue represents the false vacuum $\phi=-1$ and the true vacuum $\phi=+1$ is represented in green.}
\label{fig_phixt}
\end{figure}

Now we turn to the effect of $\epsilon$ on kink scattering.  Figs. \ref{fig_Nbxv}b-\ref{fig_Nbxv}d show the number of bounces as a function of the initial velocity for several values of $\epsilon\neq0$. They show the  inelastic scattering behavior with 1-bounce for $v>v_{crit}$, but with $v_{crit}$ growing with $\epsilon$. This is a direct effect of the tendency of the false vacuum to decay. One example of 1-bounce collision is presented in Fig. \ref{fig_phixt}a. We observed an effect that appears only for $\epsilon\neq0$, that is the lack of bounces or even bion states for $v<v^*$: the kink-antikink pair has not enought energy to encounter at the center of mass before the false vacuum decay.

 An example of this effect is shown in  Fig. \ref{fig_phixt}b. Also, 2-bounce windows are still present, but are thinner in velocity for larger values of $\epsilon$. See an example of this in  Fig. \ref{fig_phixt}c.

\begin{figure}
\includegraphics[{angle=0,width=12cm}]{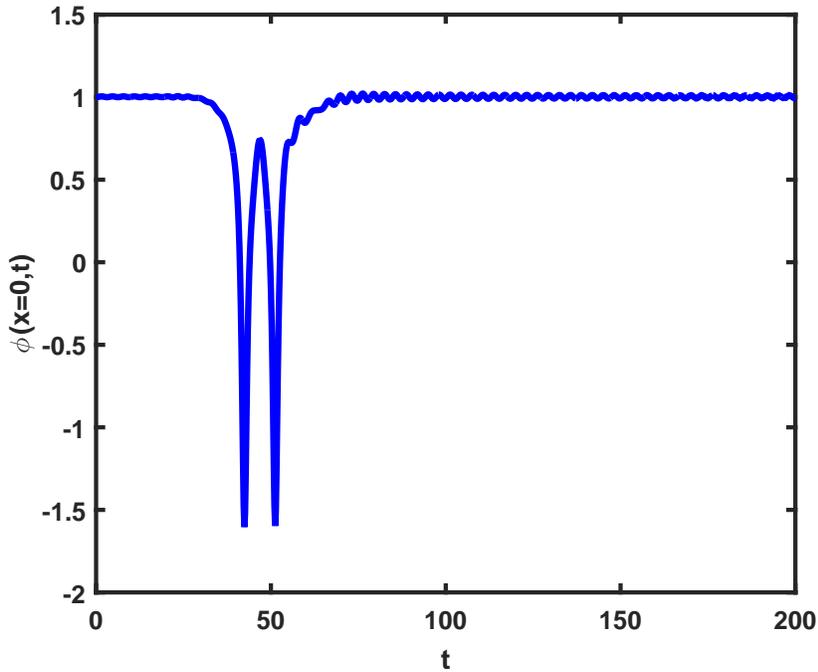}
\caption{False vacuum in an unbounded interval: kink-antikink collision showing a zero order ($m=0$) two-bounce for $\epsilon=0.01$ with $v=0.558$.}
\label{fig_bounce}
\end{figure}

Another interesting aspect of kink-antikink collisions with the growing of $\epsilon$ is the appearing of a two-bounce window of zero order for $\epsilon\gtrsim 0.008$. An example of such effect is presented in Fig. \ref{fig_bounce}. There one can see that there is no oscillations of $\phi(x=0,t)$ between the bounces, meaning that $m=0$. The appearing of zero order two-bounce windows was reported before in models with degenerate vacua in the context of boundary scattering \cite{dhmr} and in the study of transition .


\subsection{False vacuum in a bounded interval: antikink-kink collisions}
\begin{figure}
\includegraphics[{angle=0,width=8cm}]{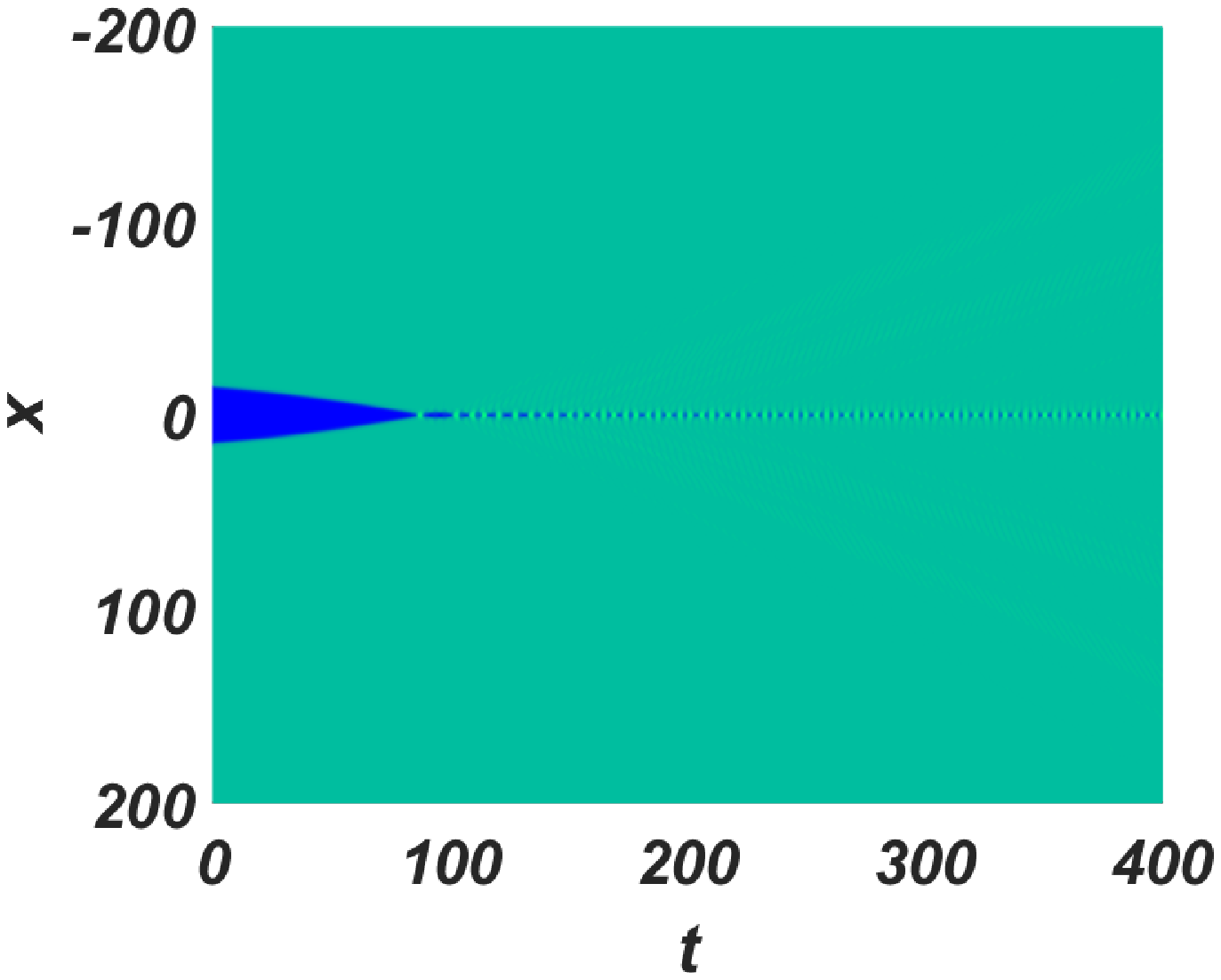} 
\includegraphics[{angle=0,width=8cm}]{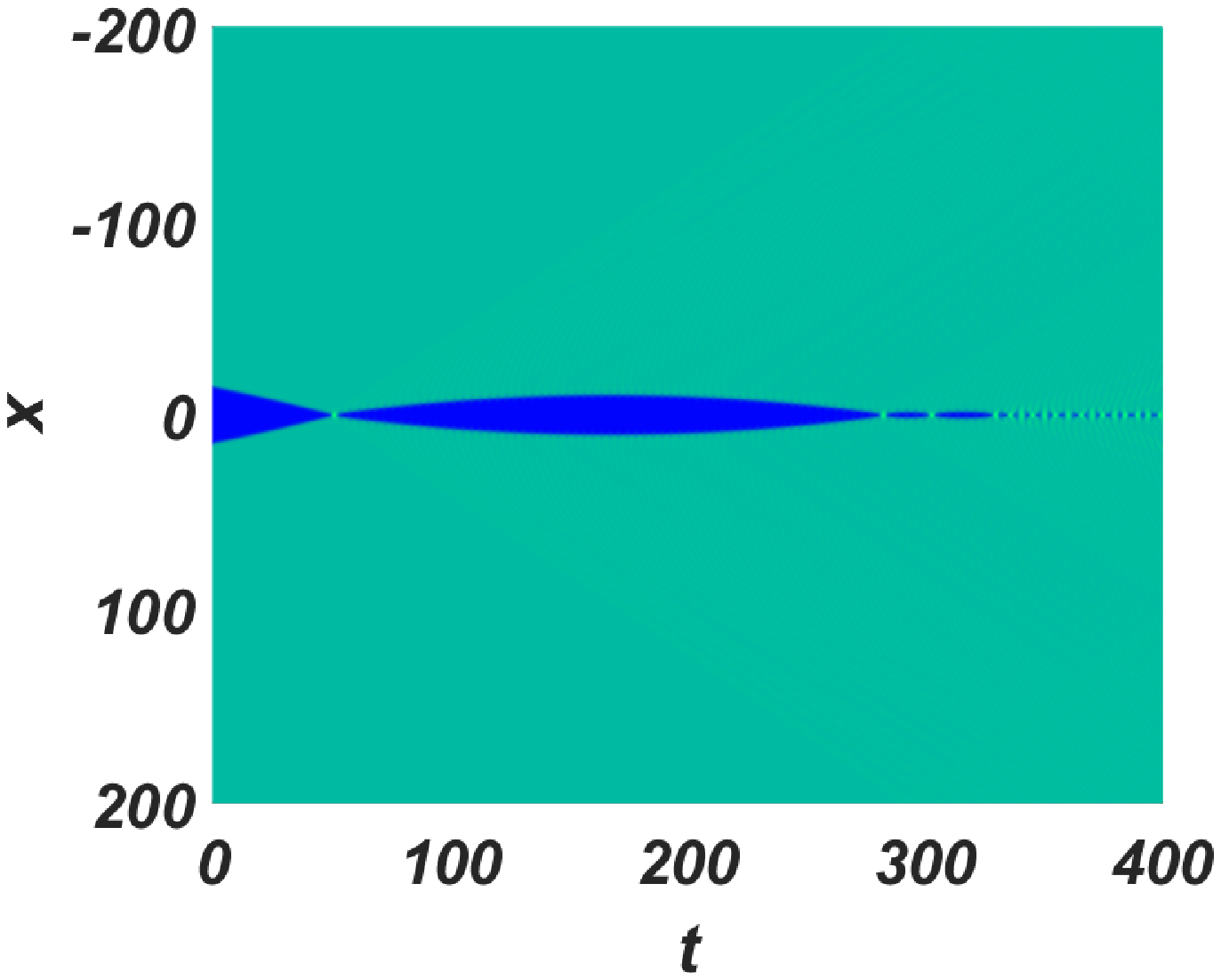} 
\includegraphics[{angle=0,width=8cm}]{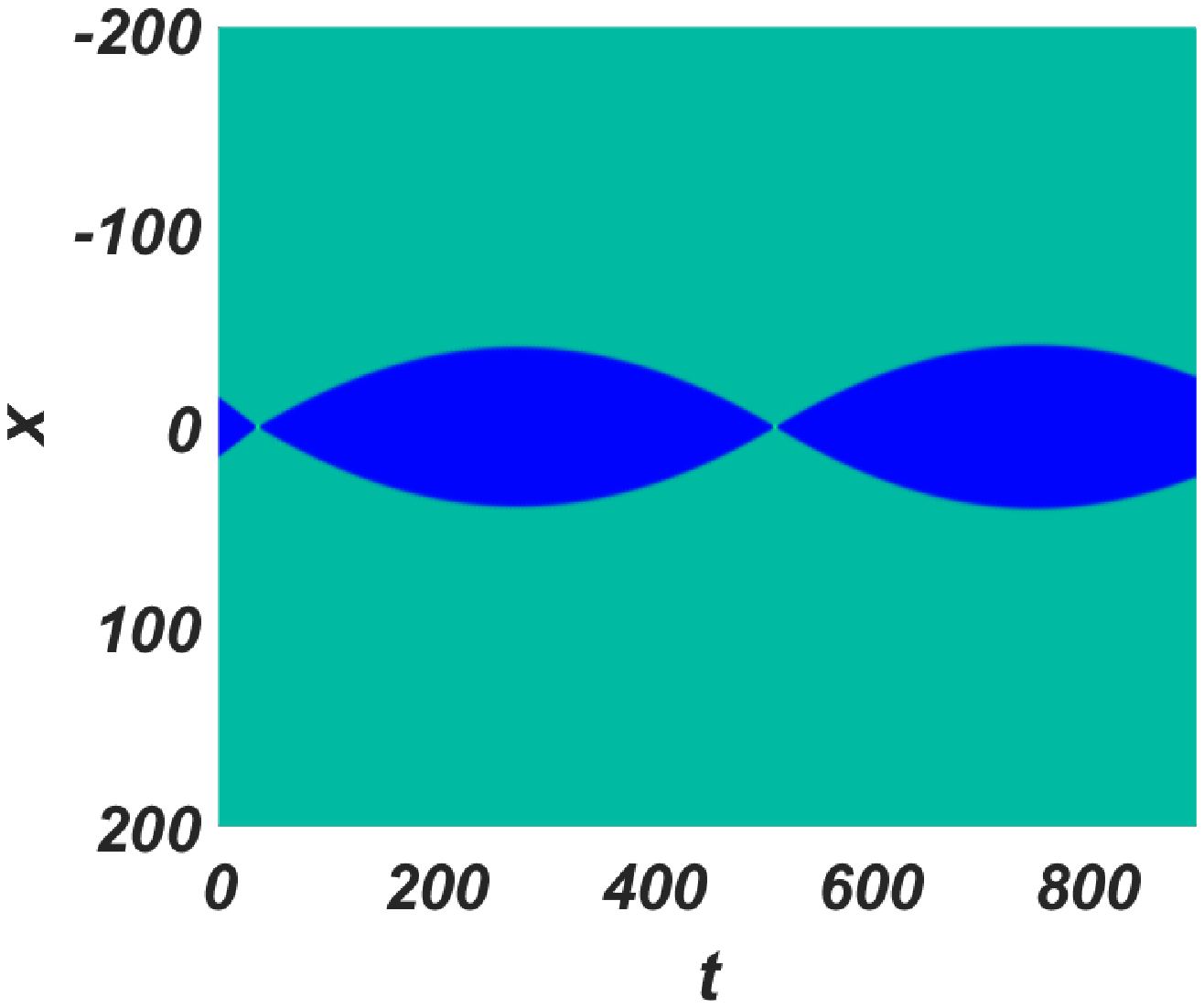} 
\includegraphics[{angle=0,width=8cm}]{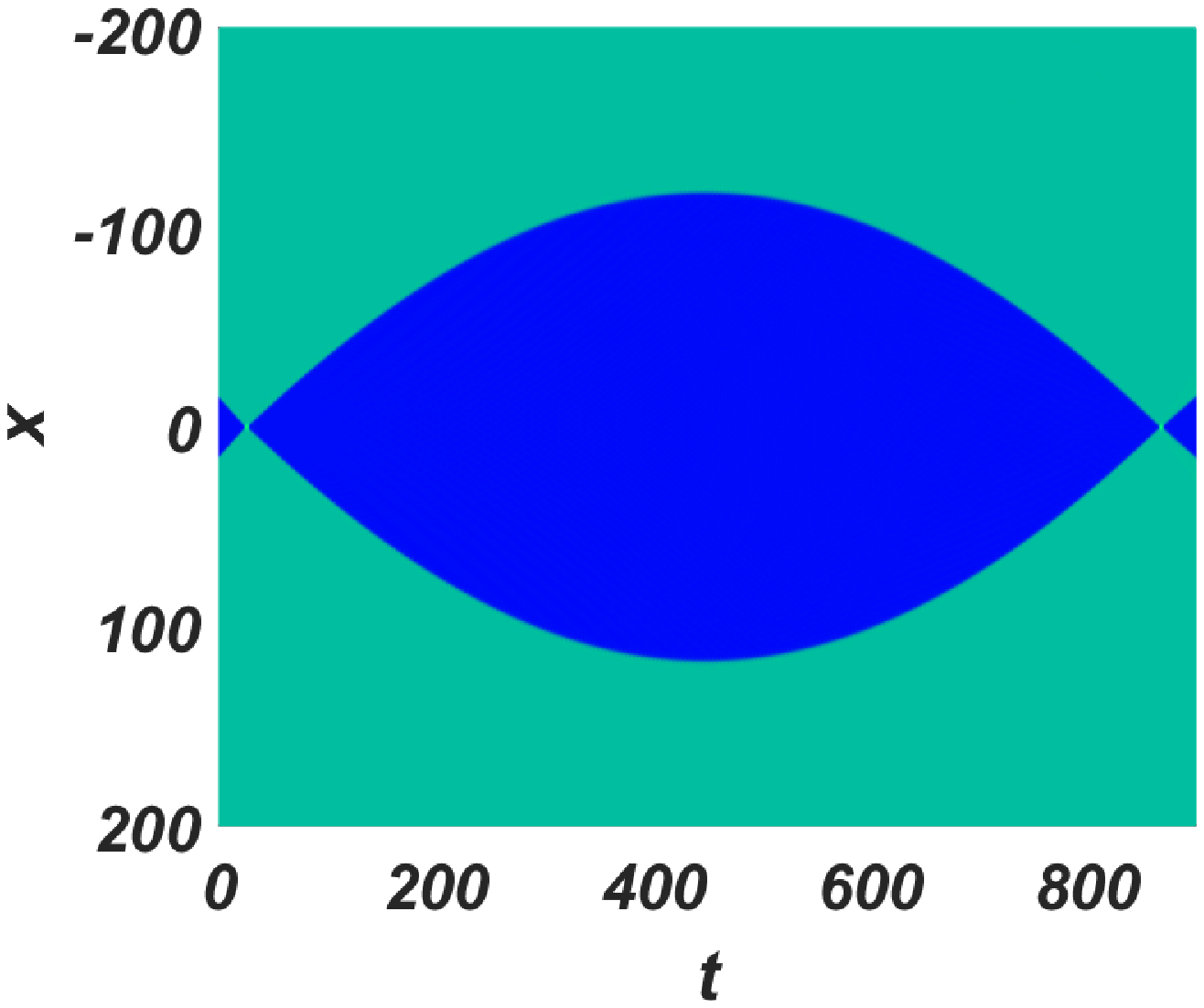}
\caption{False vacuum in a bounded interval: antikink-kink collision for $\epsilon=0.001$ and with (a) $v=0.1$ (upper left), (b) $v=0.25$ (uper right), (c) $v=0.4$ (lower left), (d) $v=0.6$ (lower right). The figures show slices in the plane $(x,t)$ of the scalar field. Blue represents the false vacuum $\phi=-1$ and the true vacuum $\phi=+1$ is represented in green.
}
\label{fig_phixt_barKK_eps0001}
\end{figure}


We take as initial condition a $\phi^4$ antikink with velocity $v$ and a $\phi^4$ kink with velocity $-v$:
\begin{eqnarray}
\phi(x,0) &=& -\phi_{K}(x+x_0,v,0)+\phi_{K}(x-x_0,-v,0)+1, \\
\dot\phi(x,0) &=& -\dot\phi_{K}(x+x_0,v,0)+\dot\phi_{{K}}(x-x_0,-v,0). 
 \end{eqnarray}
In the context of false vacuum, the output for antikink-kink collisions is much simpler than the one described above for kink-antikink. This is due to the fact that both the kink-antikink interaction and the tendency of the false vacuum to decay favor a kink-antikink pair to form.

The output of mechanism is a two-step process: firstly the field bounces around the false vacuum with low frequency. Then the antikink-kink pair radiates and the field oscillates in a high-frequency, approaching in the long run to the true vacuum state.  This is showed in Figs. \ref{fig_phixt_barKK_eps0001}a-\ref{fig_phixt_barKK_eps0001}d for $\epsilon=0.001$. Surprisingly larger initial velocities are not more effective in inducing the process of vacuum decay. On the contrary, with the growing of $v$ the low-frequency oscillation lasts even more before the decaying process to the false vacuum. It seems that, with false vacuum, larger initial velocities excite the vibrational state of the kink and antikink in a way that does not occurs for the degenerate vacuum, and the antikink-kink pair can bounce several times and be even more departed during such low-frequency oscillations (compare Figs. \ref{fig_phixt_barKK_eps0001}a-\ref{fig_phixt_barKK_eps0001}b).

\begin{figure}
\includegraphics[{angle=0,width=8cm}]{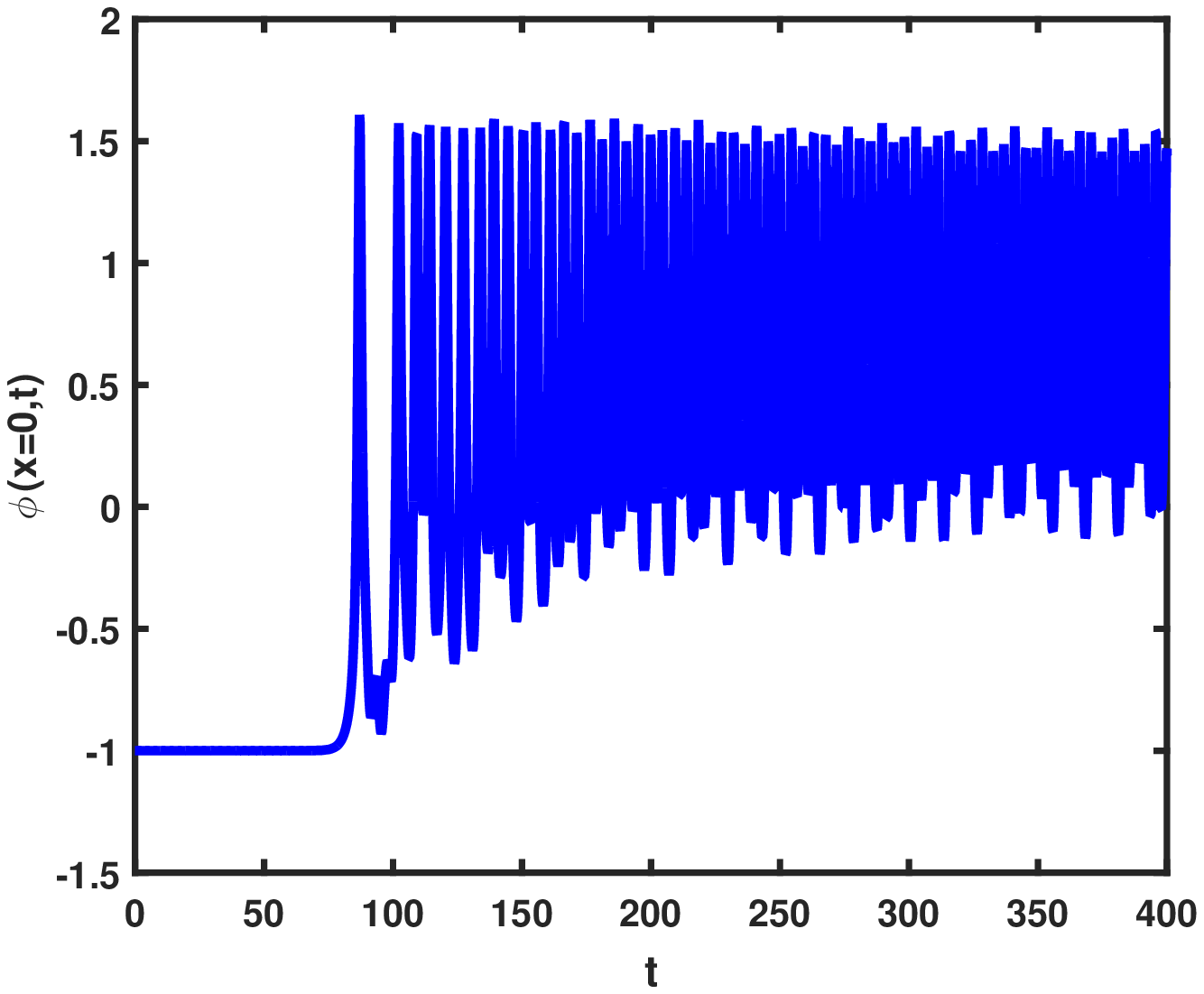} 
\includegraphics[{angle=0,width=8cm}]{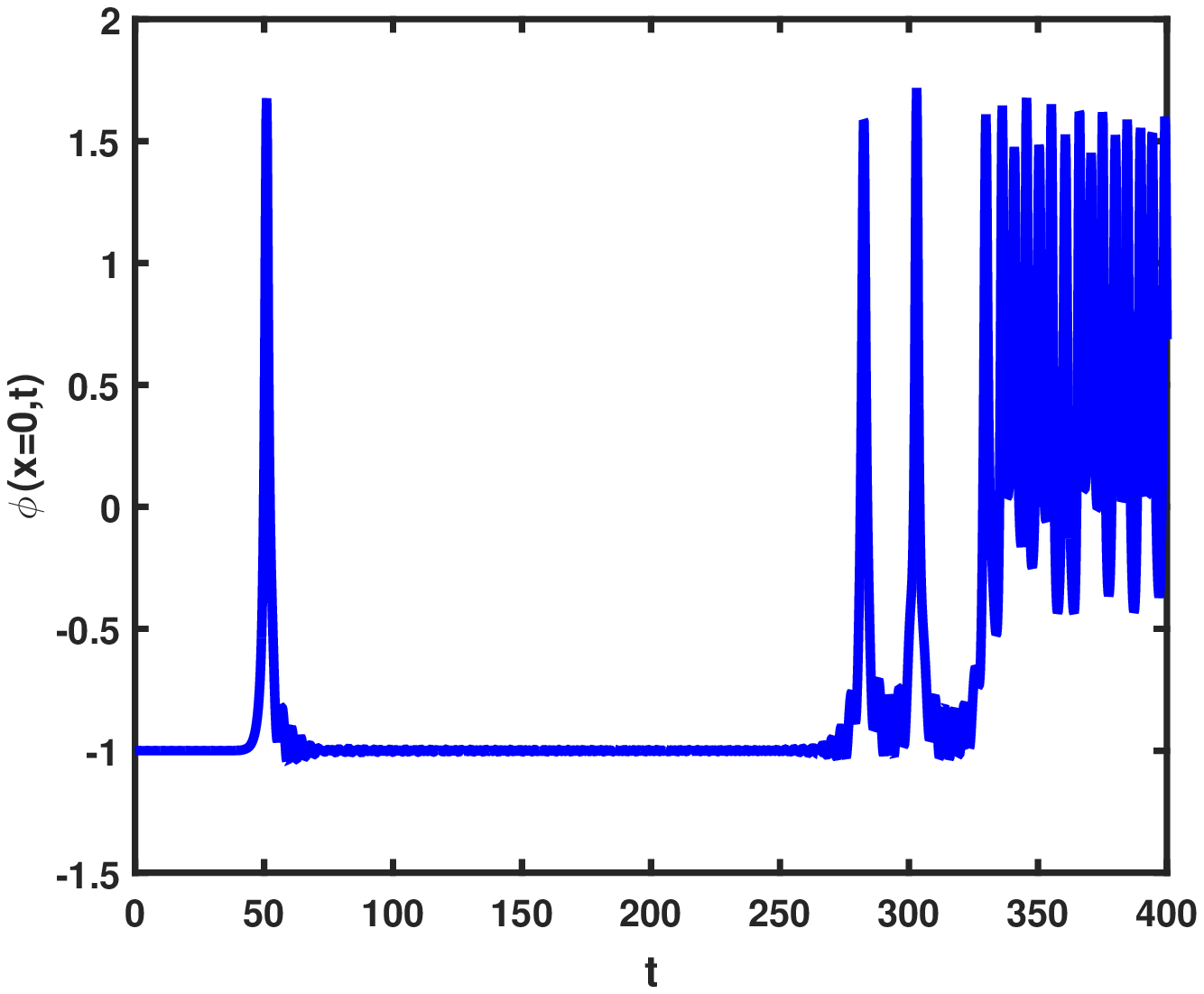}
\caption{False vacuum in a bounded interval: antikink-kink collision. Scalar field at the center of mass, $\phi(0,t)$ as a function of time for $\epsilon=0.001$ with (a) $v=0.1$ and (b) $v=0.25$, corresponding to Figs. \ref{fig_phixt_barKK_eps0001}a-b}
\label{phiCM_barKK_eps0001}
\end{figure}

 The relation between high- and low-frequency oscillations is more evident in a plot of the scalar field at the center of mass, $\phi(0,t)$, as shown in Figs. \ref{phiCM_barKK_eps0001}a-b. There one can see that for a too low initial velocity the field is not even able to bounce and the high-frequency process already takes place, as can be see in Fig. \ref{phiCM_barKK_eps0001}a for $v=0.1$. As an example for larger velocities we take $v=0.25$ as in Fig. \ref{phiCM_barKK_eps0001}b. There one can see three low-frequency oscillations around the false vacuum $\phi=-1$ before the field jumps and acquires the high-frequency oscillations that approach the true vacuum $\phi=+1$. As seen in Figs. \ref{fig_phixt_barKK_eps0001}c-\ref{fig_phixt_barKK_eps0001}d, a too larger initial velocity would require a much longer running time of simulations to see the final high-frequency decay of the false vacuum. For the values $x_0=15$ considered in the simulations, the pattern of collisions described here is valid up to $\epsilon\sim5\times 10^{-3}$. Larger values of $\epsilon$ leads to a non-recognizable pattern, meaning that the nonlinear coupling for antikink-kink collisions is already too large.


\section { Conclusions  }


In this work we have studied the effect of kink scattering in the false vacuum decay. Kink-antikink collisions are characterized by two critical velocities, $v^*$ (not present for degenerate vacua) and $v_{crit}$. For $0<v<v^*$ we have inelastic scattering of the pair without contact. For $v^*<v<v_{crit}$ the scalar field at the center of mass $\phi(x,0)$ can show bion or two-bounce around the true vacuum. In a diagram in velocities we have the usual structure of bion states and a sequence of two-bounce windows accumulating around $v_{crit}$. For $\epsilon\gtrsim \bar\epsilon$ there is the appearance of zero-order two-bounce windows. For $v>v_{crit}$ we have inelastic scattering, and $\phi(x,0)$ presents one-bounce around the true vacuum. Larger choices of initial position of the kink-antikink pair up to $x_0=25$ mean more time for the false vacuum to decay during the free propagation of the kink-antikink pair, leading to an increasing of $v^*$ and $v_{crit}$

We found that it is relatively easier for a finite region in a true vacuum to expand through the false vacuum region, as described here in kink-antikink collisions. This is evident for low initial velocities, where the kink-antikink pair scatters at non-null distance. In comparison, it demands considerably more time for a finite region in false vacuum to decay completely to the true vacuum. We showed this for antikink-kink collisions. That is, the attractive kink-antikink interaction favors the formation of a long-lived bion state that inhibits the vacuum decay. The antikink-kink collisions were described in a two-step process with a low-frequency oscillation around the false vacuum followed by a high-frequency oscillation tending to the true vacuum in the long run. This two-step process resembles the behavior of what is known in the literature as false n-bounce-windows. In the end such bounce windows must be false since the scalar field will not remain in the initial state. We also investigated the effect of increasing the gap parameter $\epsilon$.  For kink-antikink collisions we have the increasing of both velocities $v^*$ and $v_{crit}$.

For antikink-kink collisions the center of mass belongs to the false vacuum region and we have a two-step process for decay of the false vacuum: a low-frequency fast decaying bouncing state, followed by another long-lived high frequency bion state that radiates continuously. For antikink-kink collisions, the more notable effect of the increasing of $\epsilon$ is the reduction of the maximum separation of the antikink-kink pair during the low-frequency oscillations. This shows the stronger tendency of the false vacuum to decay. The final state is the same, with the field oscillating fastly around the true vacuum and radiating.


\section{Acknowledgements}
The authors thank FAPEMA and CNPq (brazilian agencies) for financial support. Gomes thanks CAUP from University of Porto for the hospitality. P.P.A. acknowledges the support provided by the FCT grant UID/FIS/04434/2013.


\end{document}